\documentclass[pra, aps, twocolumn, floatfix, showpacs]{revtex4}
\usepackage{graphicx, amsmath, amssymb, times}

\begin{document}
\title{Superfluid phases of ultracold Fermi gases on a checkerboard superlattice}
\author{M. Iskin}
\affiliation{
Department of Physics, Ko\c c University, Rumelifeneri Yolu, 34450 Sar{\i}yer, Istanbul, Turkey.
}
\date{\today}

\begin{abstract}

We analyze the ground-state phase diagram of two-component Fermi gases loaded 
into a two-dimensional checkerboard superlattice, i.e. a double-well optical lattice,
potential within the BCS mean-field theory. 
We show that, by coupling the two $s$-wave sublattice superfluid order parameters, 
a checkerboard potential gives rise to a Hamiltonian that has the form of a two-band 
superfluidity with three (two intraband and an interband) nonlocal order parameters. 
We study the evolution of these order parameters as a function of particle filling, 
interaction strength and checkerboard potential, and find that the system always 
prefers the $0$-phase solutions, i.e. the phase difference between sublattice order 
parameters is 0, but never the $\pi$-phase one. In addition, we find that the
ground-state of the system undergo a superfluid-normal quantum phase transition 
at half fillings beyond a critical checkerboard potential $C$, the threshold of which 
is precisely determined by the magnitude of the order parameter at $C = 0$, and
that the normal state rapidly turns into a checkerboard insulator as $C$ increases.

\end{abstract}
\maketitle

\section{Introduction}
\label{sec:intro}

Optical superlattices~\cite{sebby06, lee06, blochdw, blochdwgauge} 
have been of interest to the cold atom community for a long time now, and 
they have recently regained more interest in the past few years since
they may allow for the studies of topological phases of matter in the
cold-atom context~\cite{moller, dalibard, smitha, smithb}.
One way to realize a checkerboard super optical lattice is to superimpose 
two independent optical standing waves differing in period by a factor of two,
e.g. $\lambda/2$ and $\lambda$, and tunable intensities and relative phases. 
In particular, a double-well optical lattice can be produced by arranging the 
shorter-wavelength ($\lambda/2$) lattice potential in such a way to split 
the each potential well of the longer-wavelength ($\lambda$) lattice 
into two. In addition, the energy difference between the wells of the 
resultant double-well potential can also be controlled by tuning the relative 
phase of the optical potentials~\cite{sebby06, lee06, blochdw, blochdwgauge}.

In spite of serious challenges in producing fermion superfluids in earlier 
optical lattice experiments~\cite{inguscio, modugno, kohl, stoferle, bongs}, 
there is some recent experimental evidence for superfluid, metallic and 
insulating phases~\cite{mit-lattice, schneider, jordens} 
(see also the recent review~\cite{FHreview}). 
Motivated by these experiments. in this paper, we investigate the ground-state 
phases of two-component Fermi gases loaded into a two-dimensional 
checkerboard superlattice. For this purpose, we study a Fermi-Hubbard 
type lattice model which includes, in addition to the usual nearest-neighbor 
hopping and onsite (attractive) interaction, an onsite energy difference 
between sublattice sites, i.e. a staggered checkerboard potential. 
We note that the the phase diagram of the Bose-Hubbard versions of 
such a model have recently been studied for the hardcore~\cite{hencb} 
and softcore~\cite{iskincb} bosons.

Our main findings, within the single-band tight-binding BCS mean-field theory, 
are as follows.
First, we show that the $s$-wave sublattice order parameters, which are momentum 
independent in the original Hamiltonian, are coupled by the presence of
a checkerboard potential, and this gives rise to a Hamiltonian that has the 
form of a two-band superfluidity with three (two intraband and an interband) 
nonlocal (momentum-dependent higher partial waves) order parameters 
in the basis where the single-particle Hamiltonian is diagonal.
We study the evolution of these order parameters as a function of particle filling, 
interaction strength and checkerboard potential, and found that the system always 
prefers the $0$-phase solutions, i.e. the phase difference between sublattice order 
parameters is $0$, but never the $\pi$-phase one. In addition, we show that the 
ground-state of the system undergo a superfluid-normal quantum phase 
transition at half fillings beyond a critical checkerboard potential $C$, the 
threshold of which is precisely determined by the magnitude of the order 
parameter at $C = 0$, and that the normal state rapidly turns into a checkerboard 
insulator as $C$ increases.

The rest of the paper is organized as follows. We introduce the single- and
many-body Hamiltonians in Sec.~\ref{sec:ham}, and derive the 
single-quasiparticle/hole excitation spectra of the system as well as the 
complete set of self-consistency (superfluid order parameters and total 
and imbalance number) equations. We solve the resultant equations in 
Sec.~\ref{sec:numerics}, and give a detailed analysis of the obtained 
results mentioned above. 
A brief summary of our main findings is given in Sec.~\ref{sec:conclusions}.

\section{Hamiltonian}
\label{sec:ham}

It is well-established that Hubbard-type discrete lattice models can be used to 
capture the physics of cold atoms loaded into optical lattice 
potentials~\cite{jaksch, greiner02, bloch08, bloch, lin09a, lin09b}. 
For example, much of the theoretical predictions based on the simplest 
Bose-Hubbard model~\cite{fisher} have been successfully verified with 
ultracold Bose gases loaded into optical lattices. The prime examples are 
the realizations of superfluid and Mott insulator phases as well as the transition
between the two~\cite{jaksch, greiner02, bloch08, bloch, lin09a, lin09b}.  
Motivated by this success, here we study a Fermi-Hubbard type lattice model 
to analyze the physics of ultracold Fermi gases loaded into a checkerboard 
superlattice potential, as described next.

\subsection{Single-particle problem}
\label{sec:sp}

\begin{figure}[htb]
\centerline{\scalebox{0.5}{\includegraphics{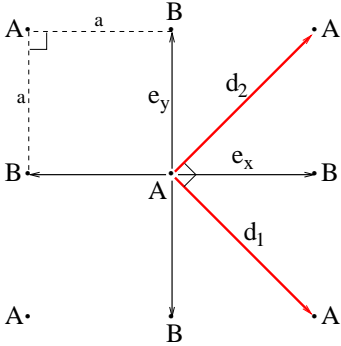}}}
\caption{\label{fig:cb} (Color online)
The checkerboard superlattice consisting of interpenetrating $A$ and $B$ square sublattices
is sketched. Here, $a$ is the lattice spacing, and 
$(\mathbf{e_x} = a\mathbf{\widehat{x}}, \mathbf{e_y} = a\mathbf{\widehat{y}})$ and 
$(\mathbf{d_1} = \mathbf{e_x} - \mathbf{e_y}, \mathbf{d_2} = \mathbf{e_x} + \mathbf{e_y})$ 
are the primitive unit vectors used to construct the single-particle problem.
}
\end{figure}

Let us first discuss the single-particle problem on a two-dimensional checkerboard 
superlattice which consists of two interpenetrating square sublattices as illustrated 
in Fig.~\ref{fig:cb}. Within the tight-binding approximation, the hopping Hamiltonian 
for such a lattice can be written as
$
H_0 = -\sum_{i \in \alpha, j \in \beta, \sigma} t_{i \alpha, j \beta, \sigma} c_{i \alpha \sigma}^\dagger c_{j \beta \sigma},
$
where $\{\alpha, \beta\} = (A, B)$ labels the sublattices and $\sigma = (\uparrow, \downarrow)$ 
labels the two components of the Fermi gas, $t_{i \alpha, j \beta, \sigma}$ is the 
tunneling (hopping) amplitude of $\sigma$ fermions between lattice sites $i \alpha$ and 
$j \beta$,  and $c_{i \alpha \sigma}^\dagger$ $(c_{i \alpha \sigma})$ operator 
creates (annihilates) a $\sigma$ fermion at lattice site $i \alpha$. In this paper, 
we limit ourselves to the simplest model where $t_{i \alpha, j \beta, \sigma} = t_{\ell \sigma}$ 
is nonzero for all nearest-neighbor lattice sites but 0 otherwise. However, we also 
allow the possibility of alternating hopping elements with a checkerboard pattern 
where $\ell$ may take two values. In particular, let's assume $t_{1 \sigma}$ 
($t_{2 \sigma}$) corresponds to the hopping amplitude from the sublattice $A$ to the 
sublattice $B$ in the positive (negative) $x$ and $y$ directions.
We note that the next-nearest-neighbor hoppings are suppressed by an order of 
magnitude compared to $t_\sigma$ in a typical atomic setting~\cite{FHreview}.

The eigenvalues of such real-space hopping Hamiltonians can be obtained by taking 
advantage of the discrete translational symmetry, and transforming them to the Fourier 
(momentum) space. Therefore, we assume a single-band description, and introduce
$
c_{i \alpha \sigma} = (1/\sqrt{M_\alpha}) \sum_{\mathbf{k_\alpha} \in \textrm{1BZ}} 
c_{\mathbf{k_\alpha} \alpha \sigma} e^{i\mathbf{k_\alpha} \cdot \mathbf{r_i}},
$
where $M_A = M_B = M/2$ is the number of sublattice sites, $\mathbf{k_\alpha}$ is the 
momentum, 1BZ is the corresponding 1st Brillouin zone, and $\mathbf{r_i}$ is the position 
of the lattice site $i$. Using this transformation, we rewrite the Hamiltonian in momentum 
space as
$
H_0 = \sum_{\mathbf{k} \sigma} \left( \epsilon_{\mathbf{k} \sigma} 
c_{\mathbf{k} A \sigma}^\dagger c_{\mathbf{k} B \sigma} + \textrm{H.c.} \right),
$
where 
\begin{align}
\label{eqn:hop}
\epsilon_{\mathbf{k} \sigma} = &-2(t_{1 \sigma} + t_{2 \sigma})  \cos(k_1 d/2) \cos(k_2 d/2) \nonumber \\
&-2i(t_{1 \sigma} - t_{2 \sigma})  \cos(k_1 d/2) \sin(k_2 d/2)
\end{align}
is in general a complex number. 
Here, $k_i = \mathbf{k} \cdot \mathbf{d_i}/d$ corresponds to the projections of the momentum 
vector along the $\mathbf{d_i}$ directions (see Fig.~\ref{fig:cb}) where $d = \sqrt{2} a$,
$c_{\mathbf{k} \alpha \sigma}^\dagger$ ($c_{\mathbf{k} \alpha \sigma}$) operator creates (annihilates)
a $\sigma$ fermion with momentum $\mathbf{k} \equiv (k_x, k_y)$ on $\alpha$ sublattice,
and $\textrm{H.c.}$ is the Hermitian conjugate. 
Thus, the single-particle dispersion relations are
$
\varepsilon_{\mathbf{k} \sigma r} = r |\epsilon_{\mathbf{k} \sigma}|,
$
where $r = (+,-)$ labels the two bands.
Note that, since the translational symmetry is doubled in real space, the original 
$\mathbf{k}$ space is halved but the number of bands is doubled, in 
such a way that this dispersion recovers the usual result, 
i.e. $\varepsilon_{\mathbf{k} \sigma} = -2t_\sigma [\cos(k_x a) + \cos(k_y a)]$,  
in the absence of a sublattice structure when $t_{1 \sigma} = t_{2 \sigma} = t_\sigma$.

In addition to the hopping part, we include an onsite checkerboard lattice potential, 
i.e. an alternating energy off-set between sublattices, which is given by
$
H_{cb} = - C \sum_{i \in \alpha, \sigma} \gamma_\alpha c_{i\alpha \sigma}^\dagger c_{i\alpha \sigma},
$
where $C \ge 0$ is its strength and $\gamma_\alpha = + 1$ ($- 1$) for $\alpha = A$ ($B$). 
This  term lowers (raises) the onsite energy of the $A$ ($B$) sublattice sites by $C$.
Using the Fourier transformation described above, we rewrite the total Hamiltonian 
$H_C = H_0 + H_{cb}$ in $\mathbf{k}$ space and diagonalize it, leading to
$
H_C = \sum_{\mathbf{k} \alpha \sigma} \varepsilon_{\mathbf{k} \sigma \alpha} b_{\mathbf{k} \alpha \sigma}^\dagger b_{\mathbf{k} \alpha \sigma},
$
where 
$
\varepsilon_{\mathbf{k} \sigma \alpha} = \gamma_\alpha \sqrt{|\epsilon_{\mathbf{k} \sigma}|^2 + C^2}
$
are the single-quasiparticle/hole dispersion relations in the presence of a checkerboard potential, 
and $b_{\mathbf{k} \alpha \sigma}^\dagger$ ($b_{\mathbf{k} \alpha \sigma}$) operator 
is the new quasiparticle creation (annihilation) operator in the transformed basis. Note that
$\varepsilon_{\mathbf{k} \sigma \alpha}$ reduces to $\varepsilon_{\mathbf{k} \sigma r}$
when $C = 0$ as expected.
We note that the $c_{\mathbf{k} \alpha \sigma}$ and $b_{\mathbf{k} \alpha \sigma}$ operators 
are related via a Bogoliubov transformation
$
c_{\mathbf{k} A \sigma} = u_{\mathbf{k} \sigma A} b_{\mathbf{k} A \sigma} + u_{\mathbf{k} \sigma B} b_{\mathbf{k} B \sigma}
$
and
$
c_{\mathbf{k} B \sigma} = v_{\mathbf{k} \sigma A} b_{\mathbf{k} A \sigma} + v_{\mathbf{k} \sigma B} b_{\mathbf{k} B \sigma},
$
where $u_{\mathbf{k} \sigma \alpha}$ and $v_{\mathbf{k} \sigma \alpha}$ are the components of the 
eigenvector that corresponds to the eigenvalue $\varepsilon_{\mathbf{k} \sigma \alpha}$. The 
eigenvectors are orthonormal in such a way that
$
u_{\mathbf{k} \sigma \alpha}/v_{\mathbf{k} \sigma \alpha} = \epsilon_{\mathbf{k} \sigma}/(C + \varepsilon_{\mathbf{k} \sigma \alpha}).
$
Having discussed the single-particle Hamiltonian, next we move on to the many-particle problem.

\subsection{Many-particle problem}
\label{sec:mp}

For the many-particle problem, the effects of local (onsite) and attractive 
interparticle density-density interactions can be taken 
into account within the BCS mean-field approximation, which is known to work well for
weak interactions at all temperatures and even for moderate interactions 
at $T = 0$~\cite{leggett, nsr, bcsbecreview}. 
For this purpose, we introduce sublattice-dependent superfluid order parameters 
$\Delta_A$ and $\Delta_B$ with $s$-wave symmetry, as defined by
$
\Delta_\alpha = -(2g/M) \sum_{i \in \alpha} \langle c_{i\alpha \downarrow} c_{i\alpha \uparrow}  \rangle
$
in real or
$
-(2g/M) \sum_{\mathbf{k}} \langle c_{\mathbf{-k} \alpha \downarrow} c_{\mathbf{k} \alpha \uparrow}  \rangle
$
in $\mathbf{k}$ space, where $g \ge 0$ is the strength of the interaction 
and $\langle \cdots \rangle$ is a thermal average. 
Therefore, the interaction contribution to the $\mathbf{k}$-space 
Hamiltonian can be written as
$
H_{mf} = (M/g) \sum_\alpha |\Delta_\alpha|^2/2 
+ \sum_{\mathbf{k} \alpha} \left( \Delta_\alpha^* c_{\mathbf{-k} \alpha \downarrow} c_{\mathbf{k} \alpha \uparrow}  + \textrm{H.c.} \right).
$
We also introduce a spin-dependent chemical potential $\mu_\sigma$ term to the Hamiltonian, i.e.
$
H_\mu = -\sum_{i \in \alpha, \sigma} \mu_\sigma c_{i\alpha \sigma}^\dagger c_{i\alpha \sigma}
$ 
in real or
$
- \sum_{\mathbf{k} \alpha \sigma} \mu_\sigma c_{\mathbf{k} \alpha \sigma}^\dagger c_{\mathbf{k} \alpha \sigma}
$ 
in $\mathbf{k}$ space, which allows us to fix the number of $\sigma$ fermions 
independently of each other.

Thus, the total many-body mean-field Hamiltonian $H = H_0 + H_{cb} + H_{mf} + H_\mu$ for the 
checkerboard superlattice can be compactly written in $\mathbf{k}$ space as,
$
H = (M/g) \sum_\alpha |\Delta_\alpha|^2/2 - 2 \sum_{\mathbf{k}} \mu_\downarrow
 + \sum_{\mathbf{k}} \psi_\mathbf{k}^\dagger \mathbf{D_k} \psi_\mathbf{k},
$
where 
$
\psi_{\mathbf{k}}^\dagger = \left( 
c_{\mathbf{k} A \uparrow}^\dagger, c_{\mathbf{k} B \uparrow}^\dagger,  c_{\mathbf{-k} B \downarrow}, c_{\mathbf{-k} A \downarrow}
\right)
$
denotes the fermionic operators collectively, and the Hamiltonian matrix $\mathbf{D_k}$ is
\begin{align}
\label{eqn:ham}
\mathbf{D_k} =  \left( \begin{array}{cccc}
-\mu_\uparrow-C & \epsilon_{\mathbf{k} \uparrow} & 0 & \Delta_A \\
 \epsilon_{\mathbf{k} \uparrow}^* & -\mu_\uparrow + C & \Delta_B & 0  \\
0 & \Delta_B^* & \mu_\downarrow - C & -\epsilon_{\mathbf{-k} \downarrow} \\
\Delta_A^* & 0 &  -\epsilon_{\mathbf{-k} \downarrow}^* & \mu_\downarrow + C
\end{array} \right).
\end{align}
In this paper, we consider equal hoppings for $\uparrow$ and $\downarrow$ fermions, 
i.e. $t_{\ell \uparrow} = t_{\ell \downarrow} = t_\ell$, leading to
$
\epsilon_{\mathbf{k} \uparrow} = \epsilon_{-\mathbf{k} \downarrow}^* = \epsilon_{\mathbf{k}}.
$
The single-quasiparticle/hole excitation spectra $E_{\mathbf{k} s}$ of the interacting 
system are given by the eigenvalues of this Hamiltonian matrix, and they can be 
compactly expressed as
\begin{align}
\label{eqn:Eks}
E_{\mathbf{k} s} &= - h + \beta_s \sqrt{\varepsilon_{\mathbf{k}}^2 + \mu^2 + \sum_\alpha \frac{|\Delta_\alpha|^2}{2} - \frac{(-1)^s}{2} \sqrt{X}}, \\
\label{eqn:X}
X &= 4|\epsilon_{\mathbf{k}}|^2 \left( |\Delta_A|^2 + |\Delta_B|^2 - 2|\Delta_A \Delta_B| \cos \Phi \right)
+ 16 \varepsilon_{\mathbf{k}}^2 \mu^2 \nonumber \\
& + \left( |\Delta_A|^2 - |\Delta_B|^2 \right) \left( |\Delta_A|^2 - |\Delta_B|^2 + 8\mu C \right),
\end{align}
where $s = (1,2,3,4)$ with $\beta_1 = \beta_2 = -\beta_3 = -\beta_4 = 1$ labels 
the eigenvalues, $h =  (\mu_\uparrow - \mu_\downarrow)/2$ is the difference and
$\mu =  (\mu_\uparrow + \mu_\downarrow)/2$ is the average chemical potential,
$\varepsilon_{\mathbf{k}} = \sqrt{|\epsilon_{\mathbf{k}}|^2 + C^2}$, and
$\Phi = \phi_A - \phi_B$ is the phase difference. Here, 
we assume $\Delta_\alpha = |\Delta_\alpha| e^{i\phi_\alpha}$. 
Note that the particle-hole symmetry of the Hamiltonian implies simultaneous 
transformation of $A \to B$ and $\mu \to -\mu$.
Note also that, after setting $\Delta_A = \Delta_B = \Delta_0$ in the $C = 0$ limit, 
this expression recovers the usual result 
$
E_{\mathbf{k} s} = -h + \beta_s \sqrt{(|\epsilon_{\mathbf{k}}|-\mu)^2 + |\Delta_0|^2},
$
which is doubly degenerate since the original $\mathbf{k}$ space is halved.

Using the quasiparticle/hole excitation spectra, we obtain the corresponding 
mean-field thermodynamic potential $\Omega$ for the total Hamiltonian $H$ as 
\begin{align}
\Omega = \frac{M}{2g} \sum_{\alpha} |\Delta_\alpha|^2  
&+ T\sum_{\mathbf{k} s} \ln\left[ \frac{1 + \tanh\left( \frac{\beta_s E_{\mathbf{k} s}}{2T} \right)}{2} \right]  \nonumber \\
& - \frac{1}{2} \sum_{\mathbf{k} s} \beta_s E_{\mathbf{k} s} - 2 \sum_{\mathbf{k}} \mu,
\end{align}
where we set the Boltzmann constant $k_B$ to unity. Following the usual procedure,
we find the lowest-energy state of the system by minimizing $\Omega$ with respect to 
the amplitudes $|\Delta_\alpha|$ and phase difference $\Phi$, leading to a set of (three) 
nonlinearly-coupled equations. In addition, we may set the total $n = n_\uparrow + n_\downarrow$ 
and imbalance $p = n_\uparrow - n_\downarrow$ number fillings using the thermodynamic 
identities $n = - (1/M) \partial \Omega / \partial \mu$ and $p = - (1/M) \partial \Omega / \partial h$,
where $0 \le n_\sigma \le 1$.
This procedure leads to a set of (five) self-consistency equations that needs to be solved
simultaneously, and four of those can be explicitly written as
\begin{align}
\label{eqn:opA}
\frac{|\Delta_A|}{g} &= \frac{1}{2M} \sum_{\mathbf{k} s} \frac{\partial E_{\mathbf{k} s}}{\partial |\Delta_A|} \tanh\left( \frac{E_{\mathbf{k} s}}{2 T} \right), \\
\label{eqn:opB}
\frac{|\Delta_B|}{g} &= \frac{1}{2M} \sum_{\mathbf{k} s} \frac{\partial E_{\mathbf{k} s}}{\partial |\Delta_B|} \tanh\left( \frac{E_{\mathbf{k} s}}{2 T} \right), \\
\label{eqn:n}
n &= \frac{1}{2M} \sum_{\mathbf{k} s} \left[ 1 + \frac{\partial E_{\mathbf{k} s}}{\partial \mu} \tanh\left( \frac{E_{\mathbf{k} s}}{2 T} \right) \right], \\
\label{eqn:p}
p &= \frac{1}{2M} \sum_{\mathbf{k} s} \frac{\partial E_{\mathbf{k} s}}{\partial h} \tanh\left( \frac{E_{\mathbf{k} s}}{2 T} \right).
\end{align}
Some of these partial derivatives are long and not particularly illuminating, 
and therefore none of them are shown. Similar to the expressions above, 
the remaining (fifth) phase-difference equation can be written as
$
0 = \sum_{\mathbf{k} s} (\partial E_{\mathbf{k} s} / \partial \Phi) \tanh[E_{\mathbf{k} s}/(2 T)],
$
and since $\partial X / \partial \Phi = 8|\epsilon_{\mathbf{k}}|^2 |\Delta_A \Delta_B| \sin \Phi$,
we immediately conclude that $\Phi$ is either $0$ or $\pi$. However, our numerical results 
suggest that the $\pi$-phase solution is never realized for the particular model Hamiltonian
that we consider in this paper. Note that these $\mathbf{k}$-space summations over the 
1BZ can be converted into the $\mathbf{k}$-space integrations via
$
\sum_{\mathbf{k}} f(k_1, k_2) \equiv [M/(8\pi^2)] \int_{-\pi}^\pi  \int_{-\pi}^\pi f(x/d,y/d) dx dy.
$

Before attacking this problem via numerical means, we would like to gain some 
physical insight. For this purpose, we rewrite the mean-field Hamiltonian $H_{mf}$
in the basis of $b_{\mathbf{k} A \sigma}$ and $b_{\mathbf{k} B \sigma}$ operators, i.e. 
the one where the single-particle checkerboard Hamiltonian is diagonal. 
Up to a constant term, the resultant four terms can be compactly written as
$
 \sum_{\mathbf{k} \alpha \beta} (F_{\mathbf{k} \alpha \beta} b_{\mathbf{-k} \alpha \downarrow} b_{\mathbf{k} \beta \uparrow} + \textrm{H.c.}),
$
where the coefficients are given by
$
F_{\mathbf{k} \alpha \beta} =   \Delta_A^* u_{\mathbf{-k} \downarrow \alpha} u_{\mathbf{k} \uparrow \beta}
                                                   + \Delta_B^* v_{\mathbf{-k} \downarrow \alpha} v_{\mathbf{k} \uparrow \beta}.
$
It is easy to show that the intraband coefficients $F_{\mathbf{k} \alpha \ne \beta}$ vanish 
when $C \to 0$ as one may expect.
This immediately reveals that $H_{mf}$ has the form of a two-band superfluidity Hamiltonian 
as long as $C \ne 0$. Note that the coefficients $F_{\mathbf{k} AA}, F_{\mathbf{k} BB}$ and
$F_{\mathbf{k} AB}$ correspond, respectively, to the $\mathbf{k}$-dependent $A$ intraband,
$B$ intraband and $AB$ interband superfluid order parameters, and all of them are spin singlet 
with even parity, i.e. $F_{\mathbf{k} \alpha \beta} = F_{\mathbf{-k} \alpha \beta}$, as one may expect.
Therefore, the starting $s$-wave sublattice order parameters $\Delta_A$ and $\Delta_B$, 
which are $\mathbf{k}$ independent in the original Hamiltonian, 
are coupled by $C \ne 0$, and this gives rise to three nonlocal ($\mathbf{k}$-dependent 
higher partial waves) order parameters in the transformed basis. 
Having derived the self-consistency equations, next we are ready to present main 
findings of this work mentioned above in the introduction.

\section{Numerical Results}
\label{sec:numerics}

In this section, we present our numerical results that are obtained by solving 
Eqs.~(\ref{eqn:opA})-(\ref{eqn:p}) for a self-consistent set of $|\Delta_A|$, $|\Delta_B|$, 
$h$ and $\mu$ as a function of given $n$, $g$ and $C$ values. 
Here, we consider only the ground states of population-balanced Fermi gases, and 
set $T = 0$ and $h = 0$. Motivated by the success of earlier theoretical works on the 
BCS-BEC crossover problem~\cite{leggett, nsr, bcsbecreview},
we emphasize that the given set of four mean-field equations (that are 
suitably generalized here to the checkerboard superlattice model) 
is expected to describe the qualitative physics well for weak interactions at all $T$, 
and even for moderate and strong interactions at $T = 0$, as long as the 
single-band tight-binding approximation remains valid. 
In order to isolate the effects of a nonzero $C$ from that of checkerboard hopping, 
let us first analyze the uniform hopping $t_1 = t_2 = t$ case.

\begin{figure}[htb]
\centerline{\scalebox{0.6}{\includegraphics{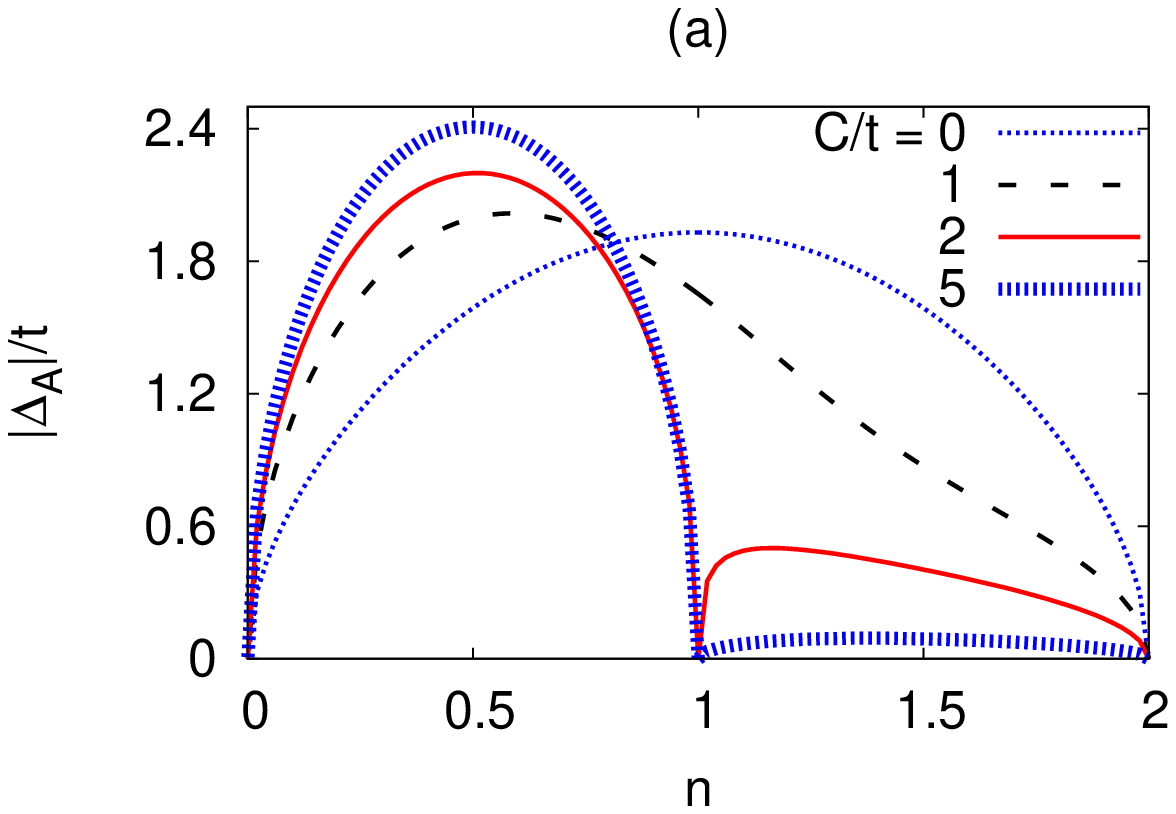}}}
\centerline{\scalebox{0.6}{\includegraphics{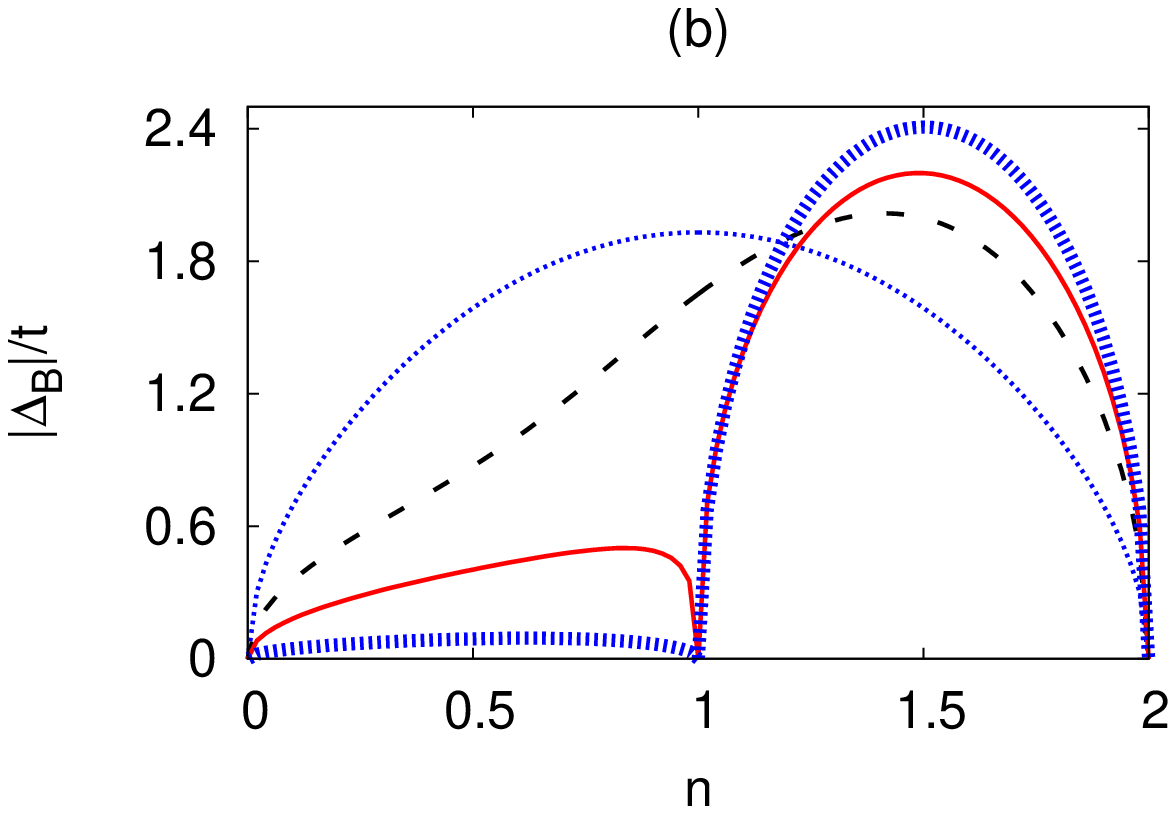}}}
\caption{\label{fig:op} (Color online)
The superfluid order parameters $|\Delta_A|$ in (a) and $|\Delta_B|$ in (b) 
(both in units of hopping $t$) 
are shown as a function of total number filling $n$ for a set of checkerboard potentials 
$C$ when the interaction strength is $g = 5t$. The particle-hole symmetry implies 
simultaneous transformation of $A \to B$ and $\mu \to -\mu$ (or $n \to 2-n)$.
}
\end{figure}
\subsection{Uniform hopping: $t_1 = t_2 = t$}
\label{sec:unhop}

\begin{figure}[htb]
\centerline{\scalebox{0.6}{\includegraphics{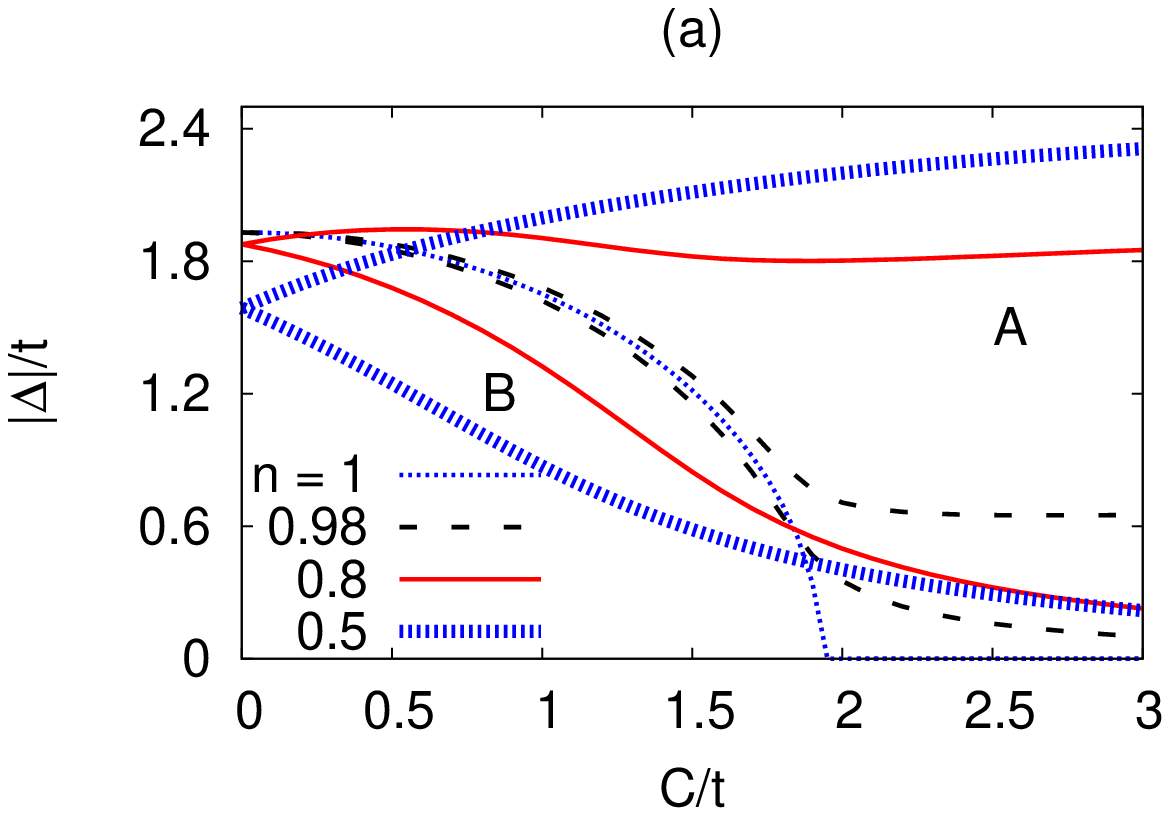}}}
\centerline{\scalebox{0.6}{\includegraphics{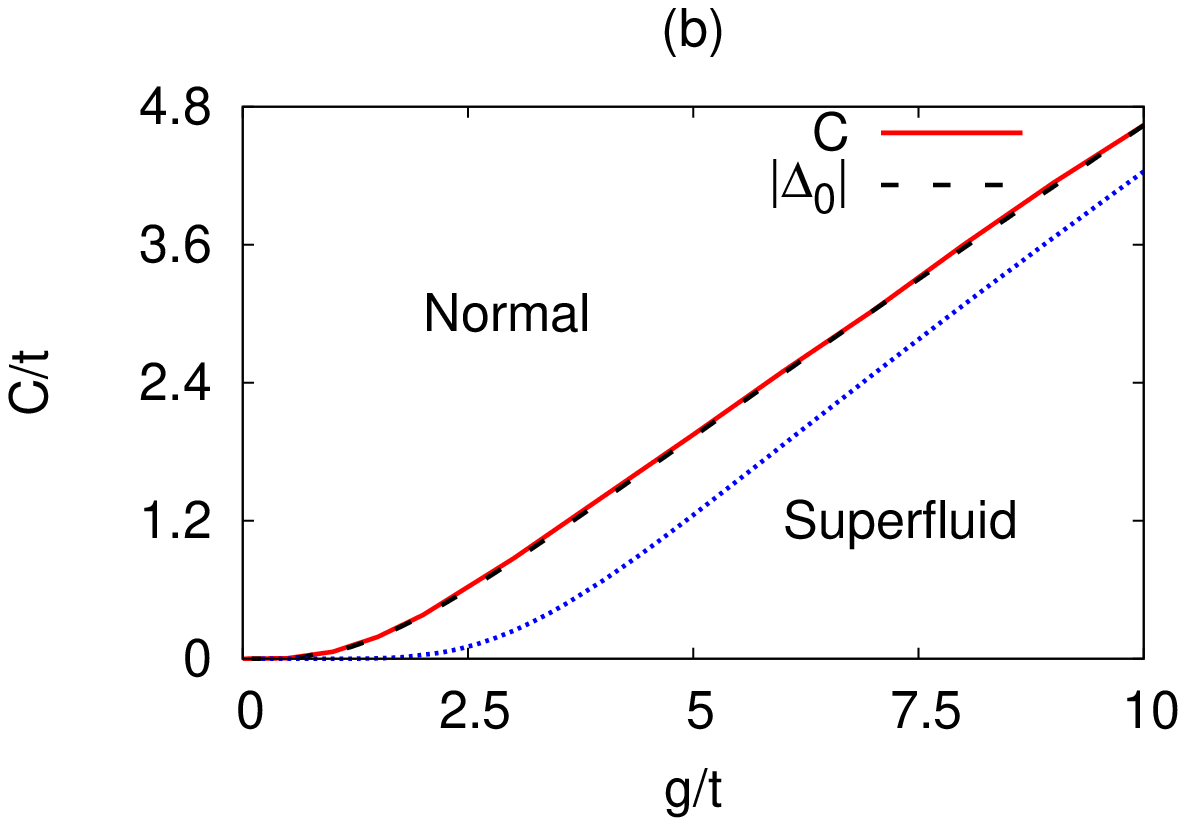}}}
\caption{\label{fig:pd} (Color online)
(a) The order parameters $|\Delta_A|$ and $|\Delta_B|$ (in units of hopping $t$) are shown 
as a function of the checkerboard potential $C$ (in units of $t$) for a set of total 
number fillings $n$ when $g = 5t$. 
(b) The superfluid-normal phase transition boundary is shown as a 
function of the interaction strength $g$ (in units of $t$) and $C$ at half filling ($n = 1$). 
Note that $C = |\Delta_0|$ coincides with the transition boundary, where $|\Delta_0|$ 
is the magnitude of the order parameter at $C = 0$. The normal state rapidly turns into 
a checkerboard insulator as $C/t \to \infty$, which is not shown in the figure but 
discussed in Sec.~\ref{sec:unhop}. See Sec.~\ref{sec:cbhop} for the blue-dotted line.
}
\end{figure}

In Fig.~\ref{fig:op}, we set $g = 5t$ and show $|\Delta_A|$ and $|\Delta_B|$ as a function 
of $n$ for a set of $C$ values. When $C = 0$, $|\Delta_A| = |\Delta_B|$ is symmetric around 
half filling ($n = 1$ or $\mu = 0$), which is a consequence of the particle-hole symmetry 
of the parent Hamiltonian, and its maximum value at half filling is a consequence of 
the lattice density of states effect. 
As mentioned in Sec.~\ref{sec:ham}, when $C \ne 0$, the particle-hole symmetry implies 
simultaneous transformation of $A \to B$ and $\mu \to -\mu$ (or $n \to 2-n)$. 
This is clearly illustrated in all of our numerical results, 
and therefore, it is sufficient to restrict our discussion only to low (particle) fillings
$0 \le  n \le 1$. We note that while $C \ne 0$ increases $|\Delta_A|$ for low fillings as a 
function of $C$, it decreases $|\Delta_B|$. This is because since $C > 0$ lowers (raises) 
the onsite energy of the $A$ ($B$) sublattice, the local chemical potential of the $A$ ($B$)
sublattice is also lowered (raised) by $C$. Therefore, the $A$ ($B$) sublattice is effectively 
becoming more and more strongly (weakly) interacting as a function of $C$. This effect 
is clearly seen in Fig.~\ref{fig:pd}(a), where we plot $|\Delta_A|$ and $|\Delta_B|$ as a
function of $C$ for a set of $n$ values. In the $C \gg t$ limit, we expect 
$|\Delta_A| = g\sqrt{n(1-n)}$ and $|\Delta_B| = 0$ for $0 \le n \le 1$, 
which is in perfect agreement with our numerical results.

In addition to these findings, we find at precisely the half filling that the system undergo 
a superfluid-normal quantum phase transition beyond a critical $C$, as illustrated 
in Fig.~\ref{fig:pd}(a). To gain intuitive understanding of this transition, we analyze the 
single quasiparticle/hole excitation spectra of the interacting system at $n = 1$, 
and therefore, set $\mu = 0$ and $|\Delta_A| = |\Delta_B| = |\Delta|$ in 
Eqs.~(\ref{eqn:Eks}) and~(\ref{eqn:X}). This gives $X = 0$ and a doubly degenerate
$
E_{\mathbf{k} s} = \beta_s \sqrt{|\varepsilon_{\mathbf{k}}|^2 + |\Delta|^2},
$
whose form is the same as the $C = 0$ spectra if we identify $|\Delta_C|^2 = |\Delta|^2 + C^2$.
Here, we recall that the single-quasiparticle/hole dispersion relation of the noninteracting 
($g = 0$) system is simply given by 
$
\varepsilon_{\mathbf{k} \alpha} = \gamma_\alpha \sqrt{|\epsilon_{\mathbf{k}}|^2 + C^2}
$
(see Sec.~\ref{sec:sp}). Therefore, these results suggest that the system does not
favor superfluidity and the normal state with $|\Delta| = 0$ becomes the ground state 
when $C \ge |\Delta_0|$, where $|\Delta_0|$ is the $C = 0$ value. 
Our numerical results shown in Fig.~\ref{fig:pd}(b) are consistent with this analysis, 
where the the critical $C$ values exactly coincide with $|\Delta_0|$ along the 
phase transition boundary. Given that 
$
|\Delta_0| = (g/2-4t^2/g) \sqrt{n(2-n)}
$
in the strong-coupling ($g \gg t$) limit, by setting $n=1$, we obtain $C = g/2-4t^2/g$ 
as the asymptotic limit of the boundary, and this is in perfect agreement with 
our numerical results.
When the critical $C \to g/2 \gg t$, we find that the $A$ ($B$) sublattice has
$n_A \to 2$ ($n_B \to 0$) so that it corresponds to a sublattice band insulator 
(fully-empty sublattice band) forming a checkerboard insulator. This intuitive 
result can also be obtained by noting that
$
n_A - n_B = -(2/M) \partial \Omega / \partial C 
= \sum_{\mathbf{k} s} (\partial E_{\mathbf{k} s} / \partial C) \tanh[(E_{\mathbf{k} s} / (2 T)],
$
which reduces to
$
n_A - n_B = (4/M) \sum_{\mathbf{k}} C/\sqrt{C^2 + |\Delta|^2 + |\epsilon_{\mathbf{k}}|^2}
$
at half-filling at zero temperature.
Therefore, the ground-state of the half-filled system first changes from a superfluid 
to normal when $C = |\Delta_0|$ at which point $|\Delta|$ vanishes, and then the normal
state rapidly turns into a checkerboard insulator as $C/t \to \infty$. 
For instance, $n_A-n_B$ becomes $1.9, 1.95, 1.99$ and $1.999$ when $C/t$ is 
approximately set to $4.9, 7.4, 17$ and $55$, respectively, in the normal state. 
Since these numbers are independent of $g/t$, the phase transition is almost 
(up to one percent deviation) directly from the superfluid to a checkerboard insulator 
when $g/t \gtrsim 34$.

We note in passing that while having a nonzero $C$ leads to a staggered pattern not only in 
the sublattice order parameters, i.e. $|\Delta_A| \ne |\Delta_B|$, but also in the sublattice 
number fillings, i.e. $n_A \ne n_B$, we avoid calling the ground-states of the system 
a supersolid (when $|\Delta_A| \ne |\Delta_B|$ and $n_A \ne n_B$) or 
a charge-density-wave insulator (when $|\Delta_A| = |\Delta_B| = 0$ and 
$n_A \ne n_B$ is an integer number). 
This is because since $C \ne 0$ breaks the translational invariance of the lattice, 
directly causing such an alternating order parameter and filling patterns, 
we believe it is important to distinguish our superfluid and checkerboard insulator 
phases from the true supersolid and charge-density-wave insulator ones, for both of 
which the translational invariance is broken spontaneously due for instance to 
the presence of nearest-neighbor interactions.

\subsection{Checkerboard hopping: $t_1 \ne t_2$}
\label{sec:cbhop}
Before we present our concluding remarks, here we discuss the possibility of having 
an alternating hopping amplitudes. Equation~(\ref{eqn:hop}) indicates that the effects 
of small deviations from the uniform hopping, i.e. when $|t_1-t_2| \ll t_1 + t_2$, can be 
taken into account (to a very good approximation) via changing the 
normalization factor $t$ that is used in the previous section to $(t_1+t_2)/2$. 
However, in the asymptotic $t_2 \to 0$ limit, Eq.~(\ref{eqn:hop}) gives
$
|\epsilon_{\mathbf{k}}| = 2 t_1 \cos(k_1 d/2),
$
and one may expect major quantitative differences. For instance, the superfluid-normal 
phase transition boundary $2C/t_1$ of a half-filled system is shown in Fig.~\ref{fig:pd}(b) 
as a function of $2g/t_1$ (blue-dotted line), and it is clearly shown that the phase 
boundary deviates substantially from that of the uniform hopping case. 
We again note that the normal state rapidly turns into a checkerboard insulator as 
$C/t_1 \to \infty$, for which $n_A-n_B$ becomes $1.9, 1.95, 1.99$ and $1.999$ 
when $C/t_1$ is approximately set to $4.2, 6.1, 14$ and $45$, respectively, 
in the normal state. Having discussed the numerical results, next we briefly 
summarize the main findings of this paper.

\section{Conclusions}
\label{sec:conclusions}

To summarize, here we studied the ground-state phases of Fermi gases 
loaded into a two-dimensional checkerboard superlattice potential, i.e. a double-well
optical lattice, consisting of two interpenetrating square sublattices $A$ and $B$. 
We described this system with a Fermi-Hubbard type lattice model which includes,
in addition to the usual nearest-neighbor hopping $t$ and onsite (attractive) 
density-density interaction $g$, a sublattice-dependent local (onsite) energy $C$. 
Within the single-band tight-binding BCS mean-field theory, we reached the 
following conclusions for such a Hamiltonian.
First, we showed that the $s$-wave sublattice order parameters 
$\Delta_A = |\Delta_A| e^{i\phi_A}$ and $\Delta_B = |\Delta_B| e^{i\phi_B}$, 
which are $\mathbf{k}$ independent in the original Hamiltonian, are coupled by 
the presence of a checkerboard potential $C \ne 0$, and this gives rise to 
a Hamiltonian that has the form of a two-band superfluidity with three 
(two intraband and an interband) nonlocal ($\mathbf{k}$-dependent higher partial waves) 
order parameters in the basis where the single-particle Hamiltonian is diagonal.
We studied the evolution of these order parameters as a function of particle filling, 
interaction strength and checkerboard potential, and found that the system always 
prefers the $0$-phase ($\phi_A = \phi_B$) solutions but never the $\pi$-phase 
($\phi_A = \phi_B + \pi$) one. 
In addition, we found at precisely half fillings that the ground-state of the system 
undergo a superfluid-normal quantum phase transition beyond a critical $C$, 
the threshold of which is precisely determined by the magnitude of the order 
parameter at $C = 0$, and that the normal state rapidly turns into a checkerboard 
insulator as $C$ increases.
One may extend this work in many ways, and motivated by the ongoing 
experiments~\cite{nistsoc, chinasocb, chinasocf, mitsoc, gaugeshaken}, 
we are especially interested in studying the effects of artificial gauge fields on 
the ground-state phases of the system, e.g. the so-called optical flux lattices.

\section{Acknowledgments}
\label{sec:ack}

This work is supported by the Marie Curie IRG Grant No. FP7-PEOPLE-IRG-2010-268239, 
T\"{U}B$\dot{\mathrm{I}}$TAK Career Grant No. 3501-110T839, and 
T\"{U}BA-GEB$\dot{\mathrm{I}}$P.

\end{document}